\documentclass[twocolumn,showpacs,preprintnumbers,amsmath,amssymb]{revtex4}

\usepackage{graphicx}
\usepackage{dcolumn}

\begin{document}
\title{Electron-positron pair production by an electron 
\\in a magnetic field in the resonant case}

\author{O. P. Novak}
\email{novak-o-p@ukr.net}
\author{R. I. Kholodov}
\email{kholodovroman@yahoo.com}

\affiliation{%
National Academy of Sciences of Ukraine, 
Institute of Applied Physics, 
58, Petropavlivska Street, 40000, Sumy, Ukraine
}

% \date{2012}

\begin{abstract}
Resonant $e^+e^-$~pair production by an electron in a magnetic field near 
the process threshold has been analytically studied. Using the Nikishov's 
theorem an estimation of the number of events has been made in the magnetic 
field equivalent to laser wave in the SLAC experiment 
[D.~Burke \textit{et al.}, Phys. Rev. Lett.~\textbf{79}, 1626 (1997)]. 
The obtained estimation is in reasonable agreement with the experimental data.
\end{abstract}

\maketitle

%==============================================================================
\section{Introduction}

Fundamental processes in intense external electromagnetic fields are 
of great interest due to the existence of strongly magnetized neutron stars 
and the construction of high-power laser systems. Known physical processes are 
modified and new ones occur in strong field environments \cite{Harding91}. 
For instance, second-order processes become more substantial, 
e.~g., double photon emission \cite{Lotstedt, Fomin}. Thus, the quantum 
electrodynamic treatment of such processes is necessary when 
field strength is comparable with the critical one 
($B_c=m^2c^3/e\hbar \approx {4.4\cdot 10^{13}}$~G).

Strong enough constant magnetic field is not feasible in laboratory 
at the present time. Nevertheless, it is possible to observe quantum 
electrodynamic (QED) processes in a strong magnetic field in experiments 
on heavy ion collisions. If the impact parameter has order of magnitude 
${\sim10^{-10}}$~sm, 
then the magnetic field of moving ions can approach magnitude of 
${\sim10^{12}}$~G in the region between the ions, while electric 
fields compensate each other. 

At the present time, FAIR (Facility for Antiproton and Ion Research) is 
under construction at the GSI Helmholtz Centre for Heavy Ion Research, 
Darmstadt, Germany. 
One of the goals of the FAIR project is to test QED in strong 
electromagnetic fields. Experiments on observation of QED processes in strong
magnetic fields in ion collisions are possible in the frame of FAIR project.

Note that the process of pair production by an electron has been 
experimentally observed in an intense laser field at SLAC National 
Accelerator Laboratory \cite{Burke}. After the SLAC experiment, 
pair creation in laser-proton collisions or in counterpropagating laser 
beams was studied in a number of works, 
e.~g. \cite{Muller03}--\cite{Muller11}. 

Electron-positron pair production 
by an electron in intense laser wave was numerically studied 
in Ref.~\cite{Hu}. In particular, the authors considered both resonant 
and nonresonant regimes of the process. In Ref.~\cite{Ilderton11} the 
trident pair production amplitude in a strong laser background 
was calculated. 

Pair production by an electron in a magnetic field was first 
studied by T.~Erber \cite{Erber66}. In Ref.~\cite{Erber66}, the rate of a cascade of 
photon emission process followed by photoproduction in a magnetic field 
has been estimated for both cases of real and virtual intermediate photon.

In the high-energy limit the considered process in arbitrary homogeneous 
constant electromagnetic field has been studied in Ref.~\cite{Baier72}.

In Ref.~\cite{Novak10} kinematics of the pair production in a magnetic field
was considered and the expressions for the total process rate containing 
integrals over orbit centers coordinates were obtained.

The purpose of the present paper is to calculate the integrals 
and obtain the explicit analytical expressions for the 
process rate. The resonant case is studied, when the rate factorizes and 
can be expressed via the product of the rates of the corresponding 
first-order processes. It is assumed that all final particles occupy 
the ground Landau level. The explicit analytical expressions for the 
total rate are obtained for subcritical  magnetic field strength, 
${B \lesssim  B_c}$.

Using Nikishov's theorem \cite{Nikishov64, FIAN} the obtained result 
has been compared with the experiment on observation of pair production 
by an electron in a laser field 
\cite{Burke}.

Relativistic units ($\hbar = c = 1$) are used throughout the paper.

%==============================================================================
\section{Process rate}

\begin{figure}
\resizebox{0.8\columnwidth}{!}{%
\includegraphics{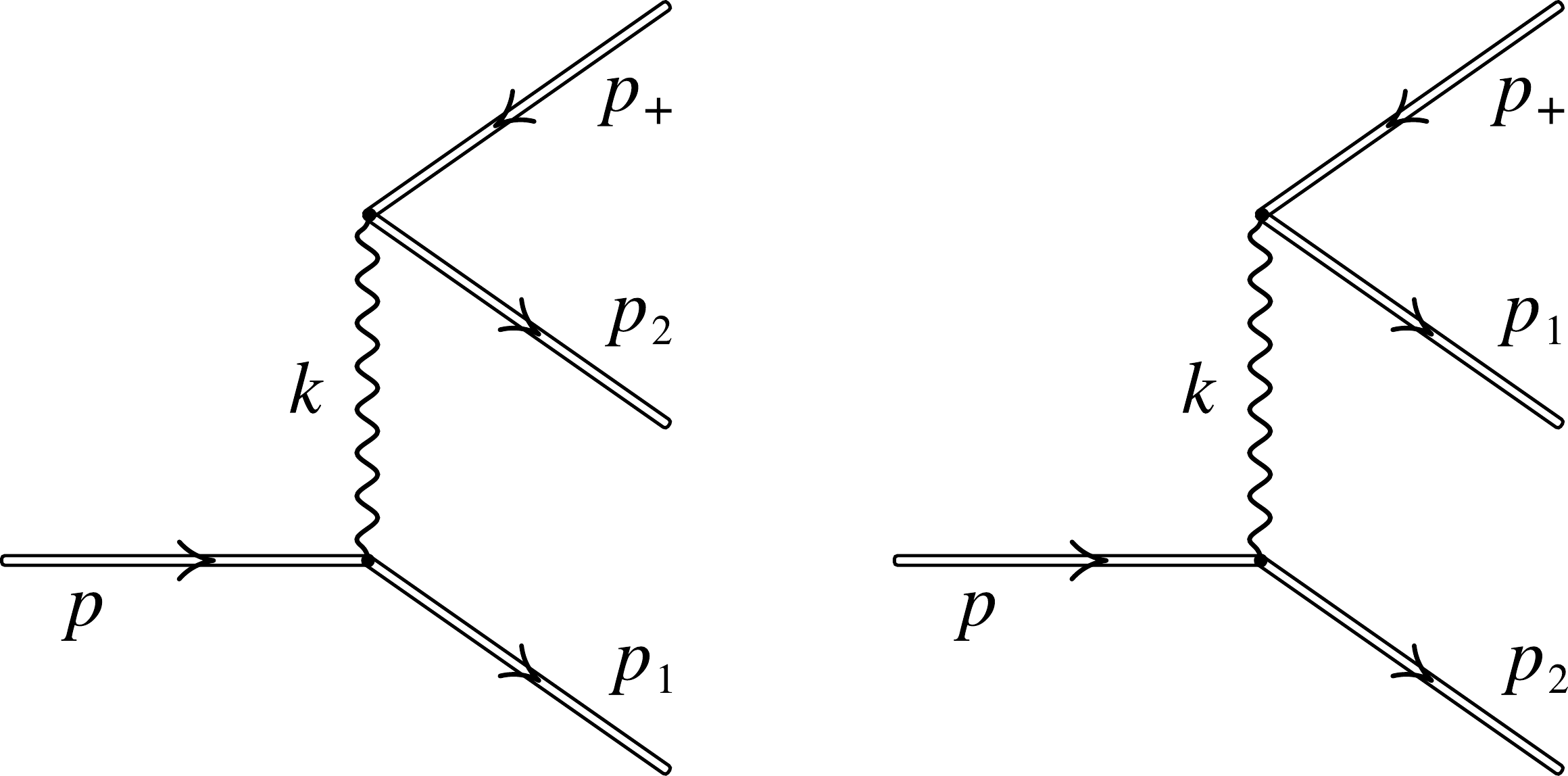}}
\caption{Feynman diagrams of the process of $e^+e^-$-pair production 
by an electron in a magnetic field.}
\label{fig1}
\end{figure}

Feynman diagrams of the considered process are shown in Fig.~\ref{fig1}, 
where the double lines represent the solutions of Dirac equations 
in a magnetic field.

The process is studied near the threshold, when the final particles occupy 
the ground Landau level. Lorentz transformation does not change magnetic 
field when passing to a reference frame moving along the field. 
Thus, without loss of generality the longitudinal momentum of the initial 
electron can be chosen equal to zero, $p_z=0$.

The  corresponding probability amplitude can be written as 
\begin{multline}
 \displaystyle
 S_{fi}=i\alpha  \iint  d^4x \, d^4x' \times\\
 \times \left[
 (\bar\Psi_2 \gamma^\mu \Psi)D_{\mu\nu}
 (\bar\Psi_1' \gamma^\nu \Psi_+') -
 \right. \\
 \left.
 -(\bar\Psi_1 \gamma^\mu \Psi)D_{\mu\nu}
 (\bar\Psi_2' \gamma^\nu \Psi_+') \right],
\end{multline}
where $\alpha$ is the fine structure constant and $D_{\mu\nu}$ 
is the photon propagator,
\begin{equation}
  D_{\mu\nu}=\frac{g_{\mu\nu}}{(2\pi)^4}
  \int d^4k\,e^{-ik(x-x')}\frac{4\pi}{k^\lambda k_\lambda},
\end{equation}
and $g_{\mu\nu}$ is the metric tensor.

The process rate is defined by the following equation:
\begin{equation}
  dW=\frac12 |S_{fi}|^2
  \frac{S d^2p_1}{(2\pi)^2}
  \frac{S d^2p_2}{(2\pi)^2}
  \frac{S d^2p_+}{(2\pi)^2}.
\end{equation}
Here, $S$ is the normalizing area, $d^2p=dp_{y}dp_{z}$.

The general expressions for the process rate look like \cite{Novak10}
\begin{gather}
\label{Wtotal+}
W^+\approx\frac{\alpha^2m}{3\pi^2\sqrt{3} l!}Y,\\
\label{Wtotal-}
W^-\sim b W^+,
\end{gather}
where the superscript denotes initial electron spin projection,  
$l$ is the Landau level number of the initial electron and
${b=B/B_c}$, $B$ is magnetic field strength and 
$B_c$ is critical field strength. The integral $Y$ has the form
\begin{equation}
\label{Y13}
Y=\iint ds\,du\left|e^{-s^2}D\right|^2,
\end{equation}
where
\begin{equation}
\label{intX}
D=\int{\frac{(s+iq)^l}{r^2-q^2}e^{-q^2-2iuq}dq}.
\end{equation}
Here, the following notations are used:
\begin{equation}
\label{ab}
\begin{array}{l}
s=m\Omega(x_0-x_{01}),\\
u=m\Omega(x_0-x_{02}),\\
q=k_x/m\sqrt{2b},\\
r^2=\Omega^2-s^2, \\
\Omega^2=2/b.
\end{array}
\end{equation}
$x_0$, $x_{01}$, and $x_{02}$ are the $x$~coordinates of the classical 
orbit centers of the initial and final electrons, respectively, and 
$k_x$ is the $x$~component of intermediate photon momentum.

The purpose of this paper is to carry out integration in 
Eqs.~(\ref{intX}) -- (\ref{Y13}) and obtain the explicit analytical 
expressions for the process rate in the resonant case. 
The integral $D$ in Eq.~(\ref{intX}) can be expressed in the form
\begin{equation}
\label{sum}
  D=\sum_{k=0}^l C_k^l s^{l-k} i^k D_k
\end{equation}
where $C_k^l=l!/k!(l-k)!$ are binomial coefficients and
\begin{equation}
  D_k=
  \int_{-\infty}^\infty q^k \frac{e^{-q^2-2iuq}}{r^2-q^2}dq.
\end{equation}
The integrand has a singularity when the condition ${r^2<0}$ is true, and 
the value of the integral (\ref{intX}) is small if $r^2$ is positive. 
Thus, it is necessary to consider the case $r^2<0$, when the inequality 
$-\Omega<s<\Omega$ is true.

The resonant divergence results in the infinite value of the process rate. 
To eliminate the divergence, one should introduce a width of the intermediate 
state $\Delta$ in accordance with Breit-Wigner prescription \cite{Graziani} 
and replace
\begin{equation}
\label{x0}
r^2\rightarrow \rho^2=r^2+ig, \qquad g=\frac{\Delta}{mb}.
\end{equation}

The integration in $D_k$ can be carried out analytically (see the Appendix) 
and the result is represented by Eq.~(\ref{app-Xkres}).

As noted in the Appendix, the quantity $D_0$ contains a divergence in the point 
$s=\Omega$. Thus, when substituting 
Eq.~(\ref{sum}) to Eq.~(\ref{Y13}), the summands with $k\geqslant 1$ 
can be neglected: 
\begin{multline} 
\label{X}
D=\frac{s^l\pi e^{-\rho^2}}{2i\rho}
\left\{ e^{-2iu\rho} \mbox{erfc}({u-i\rho})+\right.\\
\left. +e^{2iu\rho} \mbox{erfc}({-u-i\rho})\right\}.
\end{multline}

Taking into account, that the width $\Delta$ is small, 
the integration over $ds$, $du$ in Eq.~(\ref{Y13}) 
can be carried out analytically too. After the corresponding calculations 
[Eqs.~(\ref{app-intdb})--(\ref{app-Yres})] the expression $Y$ takes the form
\begin{equation}
Y=b\pi^2\sqrt{\pi}\frac{\Omega^{2l}e^{-2\Omega^2}}{\Delta/m}
\frac{\Gamma(l+1/2)}{l!},
\end{equation}
where $\Gamma(l+1/2)$ is the gamma function.

Averaging the rate over the initial electron spin projection, 
finally we obtain (in CGS units)
\begin{equation}
\label{Wppe}
W=\alpha^2\left(\frac{mc^2}{\hbar}\right)
\frac{b\sqrt{\pi}}{6\sqrt{3}}
\frac{\Omega^{2l}e^{-2\Omega^2}}{\Delta/m}
\frac{\Gamma(l+1/2)}{(l!)^2}.
\end{equation}

The quantity $\Delta$ in Eq.~(\ref{Wppe}) should be considered as the total 
width of the intermediate state. The main contribution to the width is made 
by the total radiation rate of the initial electron. There are a number of 
works related to this problem, e.~g.~\cite{Novak09}--\cite{Pavlov}.

As an example, let us calculate the rate (\ref{Wppe}) when field strength is 
$b=0.1$ ($B\approx4.4\cdot 10^{12}$~G). 
In this case the threshold Landau level number is $l=40$ and
\begin{gather}
\label{Gamma-rad}
\Delta\approx 3.9 \cdot 10^{17} \quad (\mbox{s}^{-1}),  \\
\label{Westim1}
W=1.2\cdot10^4 \quad (\mbox{s}^{-1}).
\end{gather}

\begin{figure}
\resizebox{0.8\columnwidth}{!}{%
\includegraphics{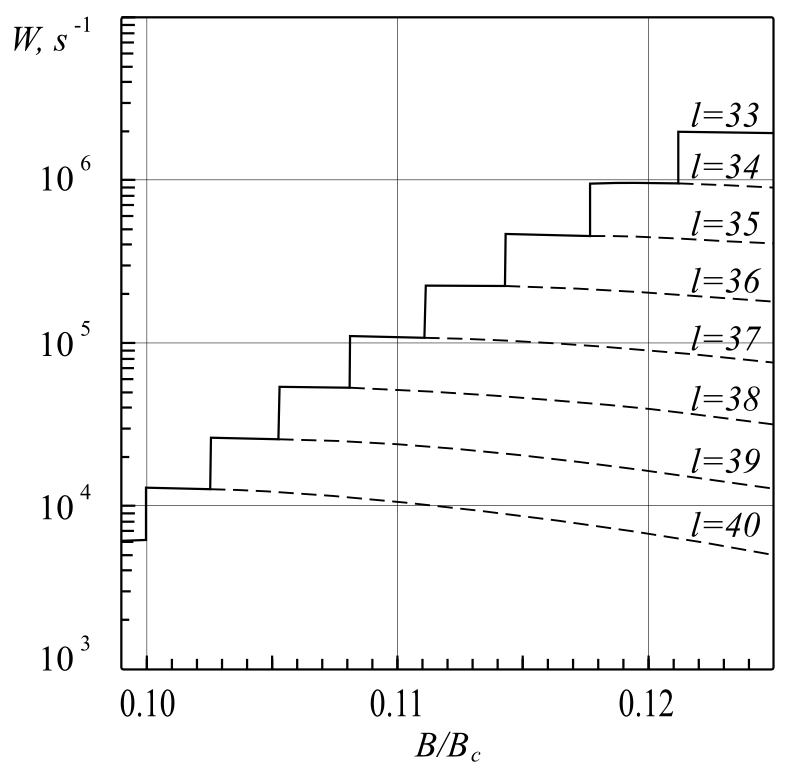}}
\caption{The dependence of the total process rate on magnetic field strength and
the initial Landau level number.}
\label{fig:rate}
\end{figure}

The dependence of the rate (14) on the magnetic field strength is shown in Fig.~\ref{fig:rate}.

%==============================================================================
\section{Factorization}

One can see that the main contribution to the process rate is made 
by the resonant mode. In this case, the total rate factorizes and 
can be expressed via the product of the rates of the first-order processes 
of magneto-bremsstrahlung and $e^+e^-$~pair production by a single photon 
\cite{Novak09, Novak08}. With account of the threshold condition 
$E\approx3m$, and consequently $bl=4$ and $l\gg1$, 
the following expression can be found:
\begin{equation}    
\label{factor}
W=\frac{\sqrt{\delta E / m}}{3\sqrt{6}}
\frac{W_{e\rightarrow\gamma e}W_{\gamma \rightarrow ee^+}}{\Delta}.
\end{equation}
Here, $W_{e\rightarrow\gamma e}$ and $W_{\gamma \rightarrow ee^+}$ 
are the rates of the corresponding firs-order processes, 
cyclotron radiation and pair photoproduction, respectively:
\begin{equation}
\label{Wrad}
W_{e\rightarrow\gamma e}=\alpha m \;\sqrt{\pi}
\frac{\Omega^{2l}e^{-\Omega^2}}{\Gamma(l+1/2)l},
\end{equation}
\begin{equation}
\label{Wpp}
W_{\gamma \rightarrow ee^+}=\alpha m
\frac{be^{-\Omega^2}}
{\sqrt{2\:\delta E/m}},
\end{equation}
where $\delta E=E-3m$ and $E = m\sqrt{1 + 2lb}$ 
is the incident electron energy.

Note that Eq.~(\ref{Wpp}) does not take account of the state widths and 
diverges if $\delta E$ goes to zero. In order that the final particles 
were not allowed to occupy exited energy levels the condition $\delta E < mb$
should be fulfilled. For exapmple, let $\delta E$ be $\frac 1 2 mb$ and 
$l = 40$, then $b = 0.10375$ and Eqs.~(\ref{Wrad}), (\ref{Wpp}) give 
the following numerical values:
\begin{gather}
W_{e\rightarrow\gamma e}= 2.1\cdot10^{13} \quad (\mbox{s}^{-1}), \\
W_{\gamma\rightarrow ee^+}=7.9\cdot10^{9} \quad (\mbox{s}^{-1}).
\end{gather}

In the review \cite{Erber66} resonant pair production by an elecron was 
considered as a cascade of synchrotron emission and photoproduction.
However, the rates of radiation and photoproduction in the high-energy 
limit were used that imply both initial and final states to be 
ultrarelativistic. This approach is not applicable near the process 
threshold when the final particles occupy the ground Landau level.

Moreover, in Ref.~\cite{Erber66} the resonant width is not taken into 
account too. In fact, decay time of the virtual state assumed to be equal 
to the half of the time of observation.

As a result, the approach of Ref.~\cite{Erber66} overestimates the process rate 
near the threshold. For the above parameters and time of observation equal 
to the doubled radiative decay time, it yields $4.7 \cdot 10^7$~s$^{-1}$ 
while Eq.~(\ref{Wppe}) gives about $1.2 \cdot 10^3$~s$^{-1}$.

%==============================================================================
\section{Discussion}

As stated in the introduction, critical or subcritical magnetic field 
is not feasible in laboratory conditions. On the other hand, 
QED processes have been already observed in SLAC experiments involving 
the interaction of an intense laser with an electron beam 
\cite{Burke, Bula, Bamber}. In Ref.~\cite{Burke} observation 
of $e^+e^-$ pair production by en electron in laser field was reported. 
About 100 positrons have been observed in 21~962 collisions of a 
${46.6}$~GeV electron beam with green ($\lambda = 527$~nm) 
terawatt laser pulses for which $\eta = 0.36$, where 
$\eta = e\sqrt{A^{\mu}A_{\mu}} / mc^2$ and 
$A_\mu$ is four-vector potential of the laser wave.

The positrons were interpreted as arising from Compton back scattering 
followed by the multiphoton Breit-Wheeler reaction,
\begin{gather}
\label{compton}
e^- +  n\omega_0 \rightarrow e^- +\omega',\\
\label{breit}
\omega' +  n'\omega_0 \rightarrow e^- +e^+,
\end{gather}
where $\omega_0$ denotes laser photons. Such a two-step process 
was distinguished from the less probable trident reaction
\begin{equation}
\label{trident}
e^- +n''\omega_0\rightarrow e^- +e^-e^+.
\end{equation}

Nevertheless, it is impossible to observe the intermediate photon 
$\omega'$ without destroying the whole process. The photon should be 
represented by an internal line in the Feynman diagram and by a photon 
propagator in the probability amplitude (but not by a wave vector). 
Consequently, to develop a consistent theory, one should consider the 
more general trident reaction Eq.~(\ref{trident}).

However, when kinematics allows on-shell intermediate state 
(so-called resonance), the Feynman diagram of the trident process 
(\ref{trident}) decomposes into two first-order diagrams corresponding 
to the processes~(\ref{compton}), (\ref{breit}). In this case the total 
rate can be expressed via the rates of the processes (\ref{compton}), 
(\ref{breit}) with some additional coefficient, that can be obtained 
only in the frame of the full theory.

It is necessary to note that Nikishov and Ritus \cite{Nikishov64} have proven 
the form of the expression of the process rate to be the same for any external 
field if the rate is expressed in terms of gauge invariants and velocity of 
the incident particle is ultrarelativistic.
In Ref.~\cite{Nikishov64} the rates of one-vertex processes were obtained in 
the case of a laser field. If the variability of the laser field is 
irrelevant, the obtained expressions reduce to the rate of the processes 
in crossed electric and magnetic fields, when 
${\vec{\mathcal{E}}  \perp \vec B}$ and $ \mathcal{E} = B$.
The total rates of such processes depends on the single invariant parameter 
$e^2(F_{\mu\nu}p_\nu)^2/m^6$ 
where $F_{\mu\nu}$ is the electromagnetic tensor and $p_\nu$ is 4-momentum.
It allows to pass to the general case of arbitrary constant field. 
In this case rates depend also on two other parameters 
$e^2F^2_{\mu\nu}/m^4$ and 
$ie^2\varepsilon_{\mu\nu\lambda\sigma}F^{\mu\nu}F^{\lambda\sigma}$
(they are equal to zero if 
$\vec{\mathcal{E}} \perp \vec B$ and $\mathcal{E} = B$).

However, since feasible fields are much less than the critical one $m^2/e$, 
these additional parameters are much less than unity.
On the other hand, if the particle energy is high enough, then these 
parameters are much less than the first one as well and could be omitted.
Therefore, the obtained rates are applicable in the case of arbitrary constant 
field, if the incident particle has relativistic energy.

In particular, considering $F_{\mu\nu}$ as a magnetic field, Nikishov and 
Ritus have obtained the results of Klepikov \cite{Klepikov} 
for intensity of a photon emission by an electron and for the rate of 
pair production by a photon in a magnetic field.

The physical reason is that due to Lorentz transformation arbitrary 
electromagnetic field goes to almost equal and almost perpendicular 
electric and magnetic fields when passing to the rest frame of the 
relativistic particle.

Thus, it is possible to compare the analytical result for the case 
of magnetic field with the experimental data of Ref.~\cite{Burke}.

If a relativistic electron propagates opposite to electromagnetic wave of 
field strength $\mathcal{E}_L$, then it experiences the field strength of 
${\mathcal{E}_0=2\gamma \mathcal{E}_L}$ in the rest frame, where $\gamma$ 
is the gamma factor. 
On the other hand, if an electron moves perpendicular to a magnetic field 
$B_{eq}$, then the field strength in the rest frame is approximately  
${\mathcal{E}_{0eq}=\gamma B_{eq}}$. Comparing $\mathcal{E}_0$ and 
$\mathcal{E}_{0eq}$ one can see that strength of the equivalent 
magnetic field in the lab frame is
\begin{equation}
B_{eq}=2\mathcal{E}_L.
\end{equation}
Note that factor 2 arises because equivalent magnetic field should take into 
account both electric and magnetic fields of the electromagnetic wave.

In order to pass to the case of alternating field of an electromagnetic 
wave, the rate $W$ (\ref{Wppe}) for the process in a magnetic field should 
be averaged over the wave period to obtain the equivalent process rate in 
laser field $W_{eq}$ \cite{Nikishov64, FIAN}:
\begin{equation}
\label{nikishov}
W_{eq}=\frac{2}{\pi}
\int\limits_0^{\pi/2}W(B_{eq}sin{\phi})d \phi.
\end{equation}
Equation~(\ref{nikishov}) allows us to compare the rates of processes in a magnetic 
field and in an intense laser wave.

However, Eq.~(\ref{Wppe}) is true near the process threshold only, when the 
condition $E\approx 3m$ is fulfilled. Therefore, it is necessary to calculate 
the rate in the moving ``threshold'' frame where the electron energy is equal 
to $E\approx 3m$, and threshold conditions are fulfilled explicitly.
The amplitude value of equal magnetic field in the threshold 
frame is $B_{eq} \approx 6.1 \cdot 10^{12}$~G and, 
consequently, $b\approx 0.14$.

It should be noted that in the SLAC experiment pair production has been 
observed near the threshold too \cite{Burke}. 
Although the electron beam energy was 46.6~GeV, the major part of this 
energy was the energy of rectilinear motion of the mass center.

It is possible to estimate the electron-laser interaction time in the 
laboratory frame $\Delta t_L$ and the number of electrons in the interaction 
region $N_{int}$ using the data from Ref.~\cite{Burke}: the electron beam size 
is $\sim 25\times40\;\mu\mbox{m}^2$, bunches contained $\sim 7\cdot10^9$ 
electrons, laser beam focal area is $30 \;\mu\mbox{m}^2$, beams crossing 
angle is $17^\circ$. Thus, 
$\Delta t_L\approx 50$~fs, 
$N_{int}\sim 2.8\cdot10^8$.

Note that to calculate the rate Eq.~(\ref{Wppe}) it is necessary to take 
into account limited interaction time as well as the radiative width 
(\ref{Gamma-rad}). Therefore, the intermediate state width is a sum of the 
radiative width and the quantity $1/\Delta t_T$ where 
$\Delta t_T=\Delta t_L/\gamma$ is laser-electron interaction time in the 
threshold frame.

The number of produced pairs can be estimated according to the expression
\begin{equation}
N_{e^+e^-}=k\cdot N_{int}(1-e^{-W_{eq} \Delta t_T}),
\end{equation}
where $k=21\:962$ is the number of collisions of the electron and laser beams 
\cite{Burke}.

The corresponding value of $\sim 80$ events is in reasonable agreement with 
the experimental result of $~106 \pm 14$ indicated in Ref.~\cite{Burke}.

Note that the authors of Ref.~\cite{Burke} pointed out 
the possible residual background of about $2\times 10^{-3}$
positrons/laser shot due to interactions of Compton
backscattered photons with beam gas. If the data are restricted
to events with $\eta > 0.216$, one can find $69 \pm 9$ positrons,
and the agreement of their number with theoretical estimations is improved.

Thus, in the present work the analytical expression for the rate of 
electron-positron pair production by an electron in a magnetic field 
near the process threshold was obtained. The number of $e^+e^-$ pairs 
created in the SLAC experiment was estimated using Nikishov's theorem. 
The obtained value is in reasonable agreement with experimental results 
as well as with numerical calculation of the Ref.~ \cite{Hu}.

We thank V.~Yu.~Storizhko and S.~P.~Roshchupkin for useful discussions.

%==============================================================================
\appendix
\section{Calculation the integrals}

To take an integral of the form
\begin{equation}
\label{app-X}
  D_0=\int\limits_{-\infty}^\infty 
  \frac{e^{-q^2-2iuq}}{\rho^2-q^2}dq
\end{equation}
it is convenient to use the apparent relation
\begin{equation}
  \int\limits_0^1 e^{t(\rho^2-q^2)}dt=\frac{e^{\rho^2-q^2}}{\rho^2-q^2}-
  \frac{1}{\rho^2-q^2}.
\end{equation}
The integral (\ref{app-X}) takes on the form
\begin{equation}
\label{app-X1}
  D_0=e^{-\rho^2}\int\limits_{-\infty}^\infty \frac{e^{-2iuq}}{\rho^2-q^2} dq +
  e^{-\rho^2}\int_0^1 \sqrt{\frac{\pi}{t}}e^{t\rho^2-\frac{u^2}{t}}dt.
\end{equation}
The first integral in (\ref{app-X1}) can be found using Jordan's lemma. 
The result is
\begin{equation}
\label{app-X11}
  \int\limits_{-\infty}^\infty \frac{e^{-2iuq}}{\rho^2-q^2} dq =
  \frac{\pi}{i\rho}e^{2i|u|q}.
\end{equation}
To find the second integral one should use the substitutions
\begin{equation}
\begin{array}{l}
    \sigma_+=\rho\sqrt{t}+{i|u|}/{\sqrt t},\\
    \sigma_-=\rho\sqrt{t}-{i|u|}/{\sqrt t}.
\end{array}
\end{equation}
After simple calculations the result of integration takes on the form
\begin{equation}
\label{app-X2}
  \frac{\pi}{2i\rho}\left[e^{-2i\rho|u|} \mbox{erfc}(|u|-i\rho)-
  e^{2i\rho|u|}\mbox{erfc}(|u|+i\rho)\right].
\end{equation}
Finally, substituting Eqs.~(\ref{app-X11}), 
(\ref{app-X2}) to Eq.~(\ref{app-X1}) the result for $D_0$ can be expressed as
\begin{equation}
\label{app-Xres}
  D_0=\frac{\pi e^{-\rho^2}}{2i\rho}
  \left[e^{-2i\rho u} \mbox{erfc}(u-i\rho)+
  e^{2i\rho u}\mbox{erfc}(-u-i\rho)\right].
\end{equation}
The above expression is valid for both $u>0$ and $u<0$ cases.

Integrals containing $q^k$ can be reduced to the considered one 
using the derivative with respect to the parameter $u$:
\begin{equation}
\label{app-Xk}
  D_k=
  \int_{-\infty}^\infty q^k \frac{e^{-q^2-2iuq}}{\rho^2-q^2}dq=
  \frac{1}{(-2i)^k} \frac{\partial^k}{\partial u^k} D_0.
\end{equation}
Taking into account the relation 
\begin{equation}
  H_n(x)=(-1)^n e^{x^2} \frac{d^n}{dx^n}e^{-x^2}
\end{equation}
where $H_n(x)$ is Hermite polynomial, the explicit form of $D_k$ 
can be expressed as
\begin{multline}
\label{app-Xkres}
  D_k=
  \frac{\pi e^{-\rho^2}}{2i\rho}\rho^k 
  \left[e^{-2iu\rho}\mbox{erfc}(u-i\rho)+ \right.\\ 
  +\left.(-1)^k e^{2iu\rho}\mbox{erfc}(-u-i\rho)\right] + \\
  +\frac{\sqrt{\pi}}{i\rho}\frac{e^{-u^2}}{(2i)^k}
  \sum_{m=1}^k C_m^k(2i\rho)^{k-m}
  \left[ H_{m-1}(u-i\rho) +\right. \\
  \left.+(-1)^k H_{m-1}(-u-i\rho) \right].
\end{multline}

Note that the value $D_0$ is in inverse proportion to $\rho$ and contains 
a divergence in the point $s=\Omega$. On the contrary, the value $D_k$ is 
finite for $k\geqslant 1$. Indeed, the summands in Eq.~(\ref{app-Xkres}) 
contain factors $\rho^{k-1}$ and $\rho^{m-k-1}$ and apparently do not 
diverge when $k > m$. When the conditions $m=k$ and $\rho=0$ are true, 
then the second summand contains the expression
\begin{equation}
  \left[ H_{k-1}(u)+(-1)^kH_{k-1}(-u)\right]=0
\end{equation}
where the relation $H_n(-x)=(-1)^nH_n(x)$ is used.

Let us proceed to calculating the integral $Y$ in Eq.~(\ref{Y13}). 
Taking into account, that the quantity $D$ in Eq.~(\ref{X}) is an even 
function of $b$, the integral over $db$ in Eq.~(\ref{Y13}) can be expressed as
\begin{equation}
\label{app-intdb}
\int_{-\infty}^{\infty}|e^{-s^2}D|^2\,du=
\frac{\pi^2 s^{2l} e^{-2\Omega^2} }{2|\rho^2|}[J_1+J_2]
\end{equation}
where 
\begin{equation}
\label{J1def}
J_1=\int_{-\infty}^\infty |e^{-2iu\rho}\mbox{erfc}(u-i\rho)|^2du,
\end{equation}
\begin{equation}
\label{J2def}
J_2=\int_{-\infty}^\infty e^{-4iru}\mbox{erfc}(u-ir)\mbox{erfc}(-u+ir) du.
\end{equation}

After integration by parts the quantity $J_1$ takes the form
\begin{gather}
\label{J1}
  J_1=\frac{1}{\sqrt{\pi}\Im{\rho}}
  \Re\left[
  e^{-2ig}j(\rho)
  \right],
  \\%
  j(\rho)=\int_{-\infty}^{\infty}e^{-(u+i\rho)^2}\mbox{erfc}(u-i\rho)du.
\end{gather}
The parameter $\rho$ can be eliminated from the argument of 
the exponent by introducing the new variable $t=u+i\rho$:
\begin{equation}
  j(\rho)=\int_{-\infty}^{\infty}e^{-t^2}\mbox{erfc}(t-2i\rho)dt.
\end{equation}
The derivative of $j(\rho)$ with respect to $\rho$ reduces to the 
Poisson integral and takes the form
\begin{equation}
  j'(\rho)=2\sqrt{2}ie^{2\rho^2}.
\end{equation}
This differential equation can be easily solved. Finally, 
after substituting the result into Eq.~(\ref{J1}) the quantity 
$J_1$ takes on the form
\begin{equation}
  J_1=\frac{\Re\left[e^{-2ig}
 \mbox{erfc}(-i\rho\sqrt{2})\right]}{\Im(\rho)}.
\end{equation}
Note that $J_1$ can be expressed as
\begin{equation}
\label{J1mod}
  J_1\approx \frac{1}{\Im (\rho)}+
  \frac{\Re[e^{-2ig}\mbox{erf}(i\rho\sqrt{2})]}{\Im (\rho)}.
\end{equation}

The integral $J_2$ in Eq.~(\ref{J2def}) can be calculated in 
the same way and looks like
\begin{equation}
  J_2=\frac{\mbox{erf}(ir\sqrt{2})}{ir}.
\end{equation}

The value of the integral over $s$ is determined by the region in the 
vicinity of the point $s=\Omega$ due to the presence of the factor 
$s^{2l}$ in the integrand in Eq.~(\ref{app-intdb}). In the points 
$s=\pm \Omega$ the first summand in $J_1$ Eq.~(\ref{J1mod}) goes to 
$\sqrt{2/g}$, while the second one and the quantity $J_2$ go to 
$\pm\sqrt{8/\pi}$. Thus, 
\begin{equation}
  Y=\pi^2e^{-2\Omega^2}
  \int_{-\Omega}^{\Omega} \frac{s^{2l}}{|\rho^2|} \frac{ds}{\Im(\rho)}.
\end{equation}
Taking into account that 
\begin{equation}
  \Im(\rho)=\frac{1}{\sqrt{2}}\sqrt{\sqrt{\rho^4+g^2}-\rho^2}
\end{equation}
and introducing a new variable $x=s/\Omega$, the quantity $Y$ 
can be transformed to
\begin{multline}
  Y=\pi^2\sqrt{2}\Omega^{2l}e^{-2\Omega^2}\frac{1}{g}\times\\
  \times
  \int_0^1 x^{2l}\sqrt{
  \frac{\sqrt{(1-x^2)^2+\delta^2}+(1-x^2)}
  {(1-x^2)^2+\delta^2}
  }dx
\end{multline}
where $\delta = g^2/\Omega$. When $\delta$ goes to zero the 
integral over $x$ in the above expression converges to
\begin{equation}
  \frac{1}{\sqrt{2}}
  \frac{\Gamma(1/2)\Gamma(l+1/2)}{\Gamma(l+1)}.
\end{equation}
Thus, 
\begin{equation}
\label{app-Yres}
  Y=
  \pi^2\sqrt{\pi}\Omega^{2l}e^{-2\Omega^2}\frac{1}{g}\frac{\Gamma(l+1/2)}{l!}.
\end{equation}

%==============================================================================

\end{document}